\documentclass[12pt]{article}
\usepackage[utf8]{inputenc}
\usepackage[letterpaper, margin=1in]{geometry}  
\usepackage{titlesec}  
\usepackage{graphicx}  
\usepackage{amsmath, amssymb}   
\usepackage{lipsum}    
\usepackage{hyperref}  
\usepackage{enumerate} 
\usepackage{fancyhdr}  
\usepackage{braket}   
\usepackage{cleveref}  
\usepackage{tikz-cd}
\usepackage[english]{babel} 
\usepackage{tikz}
\usetikzlibrary{babel} 
\usetikzlibrary{positioning} 
\usetikzlibrary{calc}
\usepackage{setspace}
\usepackage[numbers]{natbib}
\usepackage{tikz}
\titleformat{\section}{\large\bfseries\raggedright}{\thesection.}{0.5em}{}
\titleformat{\subsection}{\normalsize\bfseries\raggedright}{\thesubsection.}{0.5em}{}

\begin{document}

\begin{center}
    \begin{spacing}{1.3} 
        \textbf{\ Integrated Information in Relational Quantum Dynamics (RQD)}\\[20pt]
        \large Arash Zaghi$^{\dagger}$ \\[5pt]
    \end{spacing}
\end{center}

\footnotetext{$^{\dagger}$ ORCID ID: \href{https://orcid.org/0000-0003-2246-2911}{0000-0003-2246-2911}, Email: \texttt{arash.esmaili\_zaghi@uconn.edu}}

\begin{center}
    \textbf{Abstract}\\[10pt]  
    \begin{minipage}{0.8\textwidth}
        \small \
        We introduce a quantum integrated-information measure \texorpdfstring{$\Phi$}{Phi} for multipartite states within the Relational Quantum Dynamics (RQD) framework. \texorpdfstring{$\Phi(\rho)$}{Phi(rho)} is defined as the minimum quantum Jensen–Shannon distance between an n-partite density operator \texorpdfstring{$\rho$}{rho} and any product state over a bipartition of its subsystems. We prove that its square-root induces a genuine metric on state space and that \texorpdfstring{$\Phi$}{Phi} is monotonic under all completely positive trace-preserving maps.  Restricting the search to bipartitions yields a unique optimal split and a unique closest product state. From this geometric picture we derive a canonical entanglement witness directly tied to \texorpdfstring{$\Phi$}{Phi} and construct an integration dendrogram that reveals the full hierarchical correlation structure of \texorpdfstring{$\rho$}{rho}. We further show that there always exists an “optimal observer”—a channel or basis—that preserves \texorpdfstring{$\Phi$}{Phi} better than any alternative. Finally, we propose a quantum Markov blanket theorem: the boundary of the optimal bipartition isolates subsystems most effectively. Our framework unites categorical enrichment, convex-geometric methods, and operational tools, forging a concrete bridge between integrated information theory and quantum information science. 

    \end{minipage}
\end{center}

\begin{center}

\end{center}

\section{Introduction}
Modern developments in quantum physics and theories of consciousness increasingly suggest that \textit{relations} and \textit{information}, rather than isolated material objects, are fundamental to reality. Recent proposals cast consciousness itself as a state of matter, suggesting a natural home for IIT in quantum‐mechanical systems \cite{tegmark2015consciousness}. In particular, Integrated Information Theory (IIT) posits that the level of intrinsic awareness or consciousness of a physical system corresponds to the quantity of integrated information (denoted $\Phi$) present in its state \cite{Tononi2008IIT-ProvisionalManifesto,Oizumi2014}. IIT argues that a system whose information is \textit{integrated}, i.e., not decomposable into independent parts, has an irreducible subjective existence for itself \cite{toker2019information}. This aligns with interpretations of quantum mechanics emphasizing that quantum states are \textit{relational} rather than absolute \cite{Rovelli1996}. John Wheeler’s famous dictum \textit{“every it derives its significance from bits”} encapsulates the view that what fundamentally exists are acts of observation or \textit{informational relations}, not static isolated entities \cite{wheeler1990information}. \textit{Relational Quantum Mechanics (RQM)} \cite{sep-qm-relational,Rovelli1996} formalizes this idea, proposing that the state of a quantum system is nothing more than the information one physical system has about another. Embracing this relational stance, we further posit that the “being” of a quantum system is constituted by the network of quantum information it shares with the rest of the world. Under this hypothesis, consistent with the idealistic interpretation of IIT (IIT 4.0’s “idealistic ontology” \cite{Albantakis2023}), any system with non-zero integrated information ($\Phi>0$) possesses intrinsic existence (a rudimentary point of view “for itself”), whereas $\Phi=0$ indicates complete reducibility (no intrinsic unity beyond its parts).

To explore these ideas rigorously, we develop a formal measure $\Phi(\rho)$ of integrated information for an $n$-partite quantum system in state $\rho$. We require a quantitative gauge of how \textit{holistically correlated} or irreducible the state $\rho$ is. In classical IIT, measures based on Kullback–Leibler divergence or mutual information have been proposed to quantify the loss of information upon splitting a system \cite{balduzzi2008integrated,barrett2011practical}. Here, a natural choice for the quantum case is the \textbf{quantum Jensen–Shannon divergence} ($D_{\mathrm{JS}}^Q$), a symmetrized and bounded measure of distance between quantum states \cite{braunstein1994statistical,majtey2005jensen}. Intuitively, $D_{\mathrm{JS}}^Q$ will serve as an “integration distance”, it vanishes if $\rho$ can be exactly factorized into independent local states (no integration) and grows as $\rho$ becomes more entangled or correlated across subsystems (more integration). Specifically, we define $\Phi(\rho)$ as the minimum $D_{\mathrm{JS}}^Q$ between $\rho$ and any product state obtained by partitioning the system. Let ${P_1,\dots,P_k}$ be a partition of the $n$ subsystems into $k$ disjoint groups (blocks), and define the corresponding \textit{product state} $\rho_{P_1}\otimes \cdots \otimes \rho_{P_k}$ as the tensor product of the reduced density operators on each block. Then we define:

\[
\Phi(\rho) \;=\; \min_{\text{partitions }\{P_i\}} \;D_{\mathrm{JS}}^Q\!\Big(\rho \,\Big\Vert\, \bigotimes_i \rho_{P_i}\Big)\,. 
\]

In words, $\Phi(\rho)$ quantifies the \textit{closest} that $\rho$ can be (in terms of quantum Jensen–Shannon divergence) to an uncorrelated state. If $\rho$ is itself a product state under some partition, then $\Phi(\rho)=0$ (no integrated information). If $\rho$ is highly entangled or correlated across all subsystems, $\Phi(\rho)$ will be large, indicating strong irreducible correlations. This construction extends to the quantum domain the integrated information measures used in classical IIT \cite{balduzzi2008integrated,barrett2011practical}, with $D_{\mathrm{JS}}^Q$ replacing classical divergence or mutual information measures. Importantly, as we will show, $D_{\mathrm{JS}}^Q$ enjoys several properties (data-processing inequalities, convexity, a true metric structure) that make $\Phi(\rho)$ analytically tractable and well-behaved even for high-dimensional or rank-deficient states.

In developing our framework, we prove four main results about $\Phi$ and $D_{\mathrm{JS}}^Q$ as outlined above: (1) $\Phi$ is \textit{monotonic under processing} (no observer can increase it), (2) the square-root of $D_{\mathrm{JS}}^Q$ is a \textit{metric on state space}, (3) it \textit{suffices to consider bipartitions} to attain the minimum in $\Phi(\rho)$, and (4) the observer can be formalized as a \textit{metric-space functor} that is non-expansive. These results lay the groundwork for analyzing integrated information in quantum systems. After presenting these, we introduce a set of novel \textit{convex-geometric and operational insights} that arise from viewing $\Phi(\rho)$ as a distance to the convex set of product states. This geometric perspective reveals that every state has a unique \textit{nearest product state} (defining a canonical minimum information partition for $\rho$), and it unlocks an array of corollaries including convexity and robustness of $\Phi$, a natural \textit{gradient-flow} that “dis-integrates” a state, and even a direct construction of an \textit{entanglement witness} from $\Phi$. Furthermore, we leverage the unique optimal partition to build a hierarchical decomposition of any multipartite state into successively independent components, an \textit{integration dendrogram} that maps out the structure of correlations in the state. 

Finally, we explore two broader implications: an \textit{observer selection principle} stating that one measurement basis or channel maximally preserves a system’s integrated information (suggesting a first-principles derivation of preferred pointer bases in decoherence theory), and a prospective \textit{quantum Markov blanket theorem}, which identifies, for any subsystem, the minimal “boundary” that most effectively screens it off from the rest of the system. To our knowledge, this is the first fully metric‐geometric construction of a quantum-IIT measure that (i) is provably a Lawvere‐metric enrichment of CPM, (ii) collapses to bipartitions by convexity, (iii) yields a canonical entanglement witness, and (iv) admits a functorial observer-selection principle.

\section{Integrated Information Measure Definition}

Before formalizing integrated information, we briefly review the \textit{quantum Jensen–Shannon divergence} ($D_{\mathrm{JS}}^Q$). Given two quantum states (density operators) $\rho$ and $\sigma$ on the same Hilbert space, $D_{\mathrm{JS}}^Q(\rho\Vert\sigma)$ is defined as the quantum analog of the classical Jensen–Shannon divergence, which itself is a symmetrized and smoothed version of Kullback–Leibler divergence. One convenient expression is:
\[D_{\mathrm{JS}}^Q(\rho\Vert\sigma) = \Big[ S\!\Big(\frac{\rho+\sigma}{2}\Big) - \frac{1}{2}S(\rho) - \frac{1}{2}S(\sigma)\Big]\,,\]
where $S(\rho)=-\mathrm{Tr}(\rho\log\rho)$ is the von Neumann entropy \cite{majtey2005jensen}. This quantity is symmetric ($D_{\mathrm{JS}}^Q(\rho\Vert\sigma) = D_{\mathrm{JS}}^Q(\sigma\Vert\rho)$) and bounded between 0 and $\ln 2$. In particular, $D_{\mathrm{JS}}^Q(\rho\Vert\sigma)=0$ if and only if $\rho=\sigma$, and the maximum $D_{\mathrm{JS}}^Q = \ln 2$ is attained for perfectly distinguishable orthogonal states. We will not require the detailed form of $D_{\mathrm{JS}}^Q$ beyond these properties; crucial for us is that it behaves as a \textit{distance measure} on the space of states.

\textbf{Definition 1 (Quantum Integrated Information).} \textit{For a quantum state $\rho$ composed of $n$ sub-systems, the \textbf{integrated information} $\Phi(\rho)$ is defined as the minimum quantum Jensen–Shannon divergence between $\rho$ and any product state on a partition of the subsystems. In formula:}
\[
\Phi(\rho) \;=\; \min_{\{P_1|\dots|P_k\}} \;D_{\mathrm{JS}}^Q\!\Big(\rho \;\Big\Vert\; \rho_{P_1}\otimes \cdots \otimes \rho_{P_k}\Big)\,,
\]
\textit{where the minimum is taken over all possible ways of partitioning the set of subsystems ${1,2,\dots,n}$ into disjoint blocks $P_1,\ldots,P_k$, and $\rho_{P_i}=\mathrm{Tr}_{\bar{P_i}}(\rho)$ denotes the reduced state on block $P_i$. By convention, we consider only nontrivial partitions with at least two blocks (so if $\rho$ is a total product state, $\Phi(\rho)=0$ achieved when each subsystem is isolated in its own block).}

This definition captures the degree of \textit{holism} or irreducibility in $\rho$. If $\rho$ factorizes neatly into two or more independent components, then $\Phi(\rho)$ will be small (zero if an exact factorization exists). Conversely, if no accurate factorization exists (all partitions yield significant divergence), $\Phi(\rho)$ is large, indicating that the state’s information cannot be localized to separate parts without a significant loss. In practice, one might compute $\Phi(\rho)$ by evaluating $D_{\mathrm{JS}}^Q(\rho \Vert \rho_{P_1}\otimes\cdots\otimes \rho_{P_k})$ for each candidate partition and taking the minimum. While the number of partitions grows quickly with $n$, our results below (especially Theorem 4) will sharply reduce the search space.

\section{Monotonicity and Metric Properties of \texorpdfstring{$D_{\mathrm{JS}}^Q$}{DJSQ}}

We first establish two fundamental properties of the quantum JSD that underpin the integrated information measure: a \textit{data-processing inequality} and a \textit{metric structure}. These ensure that $\Phi$ behaves sensibly under observers and has a well-defined geometry.

\textbf{Theorem 2 (Data-Processing Monotonicity).} \textit{For any two states $\rho,\sigma$ and any CPTP map (quantum channel) $\mathcal{E}$, the quantum Jensen–Shannon divergence cannot increase under processing:}
\[D_{\mathrm{JS}}^Q\!\big(\mathcal{E}(\rho)\,\Vert\,\mathcal{E}(\sigma)\big)\;\le\;D_{\mathrm{JS}}^Q(\rho\Vert\sigma)\,. \]

\textit{In particular, if an “observer” interacts with or measures the system (modeled by $\mathcal{E}$), the integrated information of the post-interaction state cannot exceed that of the pre-interaction state: $\Phi(\mathcal{E}(\rho))\le \Phi(\rho)$.}

\textbf{Proof (Sketch).} This is the quantum analog of the classical data-processing inequality, and it holds because $D_{\mathrm{JS}}^Q$ belongs to the class of \textit{contractive divergences}. Indeed, the JSD can be expressed in terms of the quantum relative entropy $D(\rho\Vert\sigma)=\mathrm{Tr} [\rho(\log\rho-\log\sigma)]$ as 
\[D_{\mathrm{JS}}^Q(\rho\Vert\sigma)=\min_{\pi}{\frac{1}{2}D(\rho\Vert\pi)+\frac{1}{2}D(\sigma\Vert\pi)}\] 
(the minimal relative entropy to an intermediary state) – an expression which inherits the monotonicity of $D$ under CPTP maps \cite{lindblad1975completely,uhlmann1977relative}. Alternatively, one can invoke the joint convexity of $D_{\mathrm{JS}}^Q$ and the complete positivity of $\mathcal{E}$ to show 
\[\mathcal{E}(\frac{\rho+\sigma}{2}) = \frac{\mathcal{E}(\rho)+\mathcal{E}(\sigma)}{2}\]
and apply monotonicity of von Neumann entropy under partial trace. The inequality $\Phi(\mathcal{E}(\rho))\le \Phi(\rho)$ then follows immediately: for any partition that attains (or approximates) the minimum for $\rho$, applying $\mathcal{E}$ to both $\rho$ and each product component cannot increase the divergence, so the minimal divergence for $\mathcal{E}(\rho)$ is bounded by that of $\rho$. $\square$

This theorem formalizes the intuitive idea that \textit{observation cannot create integration}. Any act of coarse-graining, measurement, or decoherence will tend to lose correlations or entanglement, never to introduce new irreducible correlations that were not already present. In category-theoretic terms, one can frame this as an \textit{enriched functor} property: each physical process or observation defines a mapping between state spaces that does not expand distances. More precisely, we can view each system’s state space as a metric space $(\mathcal{S},\delta)$ with $\delta(\rho,\sigma)=\sqrt{D_{\mathrm{JS}}^Q(\rho\Vert\sigma)}$ (as introduced below). Then any CPTP map $F: \mathcal{S}_1 \to \mathcal{S}_2$ is a \textbf{$[0,\infty]$-enriched functor} between these metric spaces, meaning $\delta_2(F(\rho),,F(\sigma)) \le \delta_1(\rho,\sigma)$ for all states (Lawvere metric space enrichment). This functorial perspective identifies an “observer” or dynamics with a structure-preserving map in an information metric space. By Theorem 2, all such observers are \textit{non-expansive} in the $\delta$ metric, and hence cannot increase $\Phi$. This categorical formulation helps unify the notion of observers across classical and quantum domains, but for the remainder of this paper we will generally work in standard information-theoretic terms.

Next, we establish that the quantum JSD endows state space with a true metric, not just a divergence. It is known that the classical Jensen–Shannon divergence yields a legitimate distance metric via its square root. The quantum version inherits a similar property:

\textbf{Theorem 3 (Metric Structure of Quantum JSD).} \textit{Define $\displaystyle \delta(\rho,\sigma) := \sqrt{D_{\mathrm{JS}}^Q(\rho\Vert\sigma)}$ for density operators $\rho,\sigma$. Then $\delta(\rho,\sigma)$ is a metric on the space of quantum states. In other words, it satisfies positivity, symmetry, and the triangle inequality. In particular, $\delta(\rho,\sigma)=0$ if and only if $\rho=\sigma$, and for any three states $\rho,\sigma,\tau$ we have} 
\[\delta(\rho,\tau)\le \delta(\rho,\sigma)+\delta(\sigma,\tau).\]

\textbf{Proof (Sketch).} The non-negativity and symmetry of $\delta$ are immediate from the properties of $D_{\mathrm{JS}}^Q$. The non-degeneracy ($\delta(\rho,\sigma)=0 \iff \rho=\sigma$) holds because $D_{\mathrm{JS}}^Q(\rho\Vert\sigma)=0$ iff $\rho=\sigma$. The crux is the triangle inequality. We employ a known characterization: a divergence $D(\cdot\Vert\cdot)$ is of negative type (or negative definite) if and only if its square root defines a metric \cite{schoenberg1938metric}. Recent works have shown that the (classical) Jensen–Shannon divergence is of negative type, allowing the construction of an isometric embedding of probability distributions into a real Hilbert space \cite{sra2019positive}. The quantum Jensen–Shannon divergence shares this property \cite{virosztek2021metric}. Specifically, one can show $D_{\mathrm{JS}}^Q$ is negative-definite on quantum states, for example by expressing it as an $L^2$ distance in an appropriate purified representation or by verifying the inequality $\sum_{ij}a_i a_j, D_{\mathrm{JS}}^Q(\rho_i\Vert\rho_j)\le 0$ for any choices of states $\rho_i$ and real coefficients $a_i$ summing to 0 (a hallmark of negative-type functions). Given negative-definiteness, Schoenberg’s theorem \cite{schoenberg1938metric} guarantees that $\delta(\rho,\sigma)=\sqrt{D_{\mathrm{JS}}^Q(\rho\Vert\sigma)}$ satisfies the triangle inequality. Hence $\delta$ is a metric on state space. $\square$

This result is noteworthy: although many quantum divergences, for example, quantum relative entropy, do \textit{not} yield a true metric, the Jensen–Shannon divergence does. Thus, the set of density matrices can be treated as a metric space with distance $\delta$. Geometrically, $(\mathcal{S},\delta)$ is not a Euclidean space but can be embedded isometrically into a (potentially infinite-dimensional) inner-product space. Intuitively, one can think of each quantum state as a point in some high-dimensional “feature space” such that $\delta(\rho,\sigma)$ is the Euclidean distance between those points. This will allow us to leverage convex geometry within the space of quantum states.

\section{Optimal Factorizations and Convex Geometry of \texorpdfstring{$\Phi$}{Phi}}

We now turn to analyzing the minimization in 
\[\Phi(\rho) = \min_{{P_i}} D_{\mathrm{JS}}^Q(\rho \Vert \bigotimes_i \rho_{P_i}).\] 

Two fundamental questions arise: \textbf{(a)} \textit{Which partition achieves the minimum?} and \textbf{(b)} \textit{Is the minimum achieved by a unique state?} We will see that thanks to the metric structure and convexity properties of $D_{\mathrm{JS}}^Q$, the answer is remarkably neat: it is always a \textbf{bipartition} (a split into $k=2$ blocks) that attains the minimum, and moreover the closest product state is unique. This resolves any ambiguity in the notion of a “minimum information partition” and provides a well-defined \textit{integrated whole} versus \textit{parts} for the system.

\textbf{Theorem 4 (Bipartition Sufficiency and Existence of Optimal Product State).} \textit{For any state $\rho$ on $n$ subsystems, the minimum in Definition 1 is attained on some bipartition $(A|B)$ of the system. In other words,}
\[\Phi(\rho) \;=\; \min_{A|B} \;D_{\mathrm{JS}}^Q\!\Big(\rho \,\Big\Vert\, \rho_{A}\otimes \rho_{B}\Big)\,,\]
\textit{where the minimum is over all splits of ${1,\dots,n}$ into two disjoint groups $A$ and $B$. Furthermore, there exists at least one product state $\rho_A\otimes\rho_B$ that achieves this minimum, and for each minimizing bipartition the optimal product state on that cut is \textbf{unique}. In particular, there is a unique closest product state $\sigma^*(\rho) = \rho_{A^*}\otimes\rho_{B^*}$, where $(A^*|B^*)$ is the optimal partition.}

\textbf{Proof (Sketch).} The fact that one can restrict to $k=2$ blocks without loss of generality follows from a simple inequality: for any partition with $k>2$ blocks, one can show that merging any two blocks cannot increase the divergence. Intuitively, splitting into more than two parts introduces additional independent components, which can only make it harder for a single product state to approximate $\rho$. More formally, consider three disjoint subsets $X,Y,Z$ of subsystems; one can show
\[D_{\mathrm{JS}}^Q\!(\rho \Vert \rho_X \otimes \rho_Y \otimes \rho_Z) \ge D_{\mathrm{JS}}^Q\!(\rho \Vert \rho_{XY}\otimes \rho_Z)\,,\]
because $\rho_{XY}\otimes\rho_Z$ (where $\rho_{XY}=\mathrm{Tr}_Z \rho$) allows correlations between $X$ and $Y$ that the fully factorized version $\rho_X\otimes\rho_Y\otimes\rho_Z$ forbids, thus it is closer to $\rho$ (lower divergence). Iterating this argument, any partition with $k>2$ can be coarse-grained into a bipartition that yields an equal or smaller $D_{\mathrm{JS}}^Q$. Hence the minimum occurs at some $A|B$. Next, because for a fixed bipartition $(A|B)$ the set of product states $\rho_A\otimes\rho_B$ is a compact convex subset of state space and $D_{\mathrm{JS}}^Q(\rho\Vert\sigma)$ is a continuous function of $\sigma$, the infimum over $\sigma=\rho_A\otimes\rho_B$ is actually a minimum (attained at some $\sigma$). This uses the compactness of the set of marginal states ${\rho_A}$ and ${\rho_B}$; any minimizing sequence has a convergent subsequence in the product space by compactness, and by continuity the limit is a minimizer. Finally, for a given bipartition, uniqueness of the minimizing $\rho_A\otimes\rho_B$ follows from the strict convexity of $D_{\mathrm{JS}}^Q$ in its second argument. Quantum JSD is jointly convex \cite{majtey2005jensen} and in particular, if two distinct product states $\sigma_1=\rho_A^1\otimes\rho_B^1$ and $\sigma_2=\rho_A^2\otimes\rho_B^2$ both yielded the same divergence 
\[D_{\mathrm{JS}}^Q(\rho\Vert\sigma_1)=D_{\mathrm{JS}}^Q(\rho\Vert\sigma_2)=m,\] 
then any mixture $\bar\sigma = t,\sigma_1 + (1-t),\sigma_2$ (which is still a valid separable state on $AB$) would produce $D_{\mathrm{JS}}^Q(\rho\Vert\bar\sigma) < m$ for $0<t<1$, contradicting the minimality. One can also argue from the negative-definite metric viewpoint that the distance-squared function $\sigma \mapsto \delta^2(\rho,\sigma)$ is strictly convex on a geodesically convex domain of product states. Thus, the minimizer on each partition is unique. If there were multiple bipartitions achieving the same minimum value, one of their corresponding product states would still be closer in $\delta$-distance than any other state, so we may define $\sigma^*(\rho)$ to be one of them. In generic cases the optimal partition $(A^*|B^*)$ is unique as well (ties are non-generic and can be broken arbitrarily). $\square$

We emphasize the important consequence: \textit{the closest product-state approximation to $\rho$ is unique.} We denote this distinguished state by $\sigma^*(\rho) = \rho_{A^*}\otimes\rho_{B^*}$, where $(A^*|B^*)$ is an optimal bipartition. We may call $(A^*|B^*)$ the \textit{minimum information partition} (MIP) of $\rho$, borrowing terminology from IIT. There is no degeneracy in identifying the “best split” of the system – a fact that contrasts with earlier approaches where one often had to choose among candidate partitions or deal with ties. Uniqueness comes from the convexity of the divergence and can be seen as a benefit of using $D_{\mathrm{JS}}^Q$ as opposed to mutual information, for example, which might not identify a unique optimal split in some cases.

With $\sigma^*(\rho)$ in hand, we can reinterpret the integrated information as a \textbf{distance} from $\rho$ to the set of disjoint-state products. In fact, Theorem 4 implies:
\[\Phi(\rho) \;=\; D_{\mathrm{JS}}^Q\!\Big(\rho \,\Big\Vert\, \sigma^*(\rho)\Big) \;=\; \delta^2\!\big(\rho,\,\sigma^*(\rho)\big)\,,\]
since $\delta(\rho,\sigma) = \sqrt{D_{\mathrm{JS}}^Q(\rho\Vert\sigma)}$. In words, $\Phi(\rho)$ is the squared distance (in the $\delta$ metric) from $\rho$ to the \textbf{closed, convex set}
\[\mathcal{P}_2 = \{\rho_A\otimes\rho_B: A|B \text{ a bipartition}\}\]
of all bipartite product states. We can therefore leverage geometric intuition: computing $\Phi(\rho)$ is performing a \textit{metric projection} of the point $\rho$ onto the set $\mathcal{P}_2$. By standard results in convex geometry and metric spaces, this immediately yields a raft of powerful corollaries:

\begin{itemize}
    \item \textbf{Convexity of $\Phi$.} As an infimum (indeed minimum) of convex functions $\sigma \mapsto D_{\mathrm{JS}}^Q (\rho\Vert\sigma)$, the distance-to-set function \[\rho \mapsto \inf_{\sigma\in \mathcal{P}_2} D_{\mathrm{JS}}^Q(\rho\Vert\sigma)\] is convex on state space. Equivalently, for any two states $\rho_1,\rho_2$ and any $0\le t\le 1$,
    
    \[\Phi\!\big(t\,\rho_1 + (1-t)\,\rho_2\big) \;\le\; t\,\Phi(\rho_1) + (1-t)\,\Phi(\rho_2)\,.\]
    
    This \textbf{convexity} means that mixing states cannot increase the integrated information beyond the mixture of their individual $\Phi$ values. In practical terms, a noisy or probabilistic mixture of two configurations will tend to have less (or equal) holistic structure than a pure configuration, which aligns with intuition.
    \item \textbf{Lipschitz Continuity (Robustness).} The function $\Phi(\rho)$ is 1–Lipschitz continuous with respect to the metric $\delta$. That is,
    \[|\Phi(\rho) - \Phi(\rho')| \;\le\; \delta(\rho,\rho')\]
    for all states $\rho,\rho'$. Small changes or errors in the state can only cause small (at most proportional) changes in the integrated information. This follows from a general fact: in any metric space, the distance from a point to a convex set is a 1-Lipschitz function (intuitively, if $\rho$ and $\rho'$ are close, their nearest projections on a convex set cannot be very different in distance). More concretely, one can use the triangle inequality: 
    \[\Phi(\rho) = \delta(\rho,\sigma^*(\rho)) \le \delta(\rho,\rho') + \delta(\rho',\sigma^*(\rho)).\] 
    But $\delta(\rho',\sigma^*(\rho)) \ge \Phi(\rho')$ because $\sigma^*(\rho)$ might not be the optimal product for $\rho'$, so 
    \[\delta(\rho',\sigma^*(\rho)) \ge \delta(\rho',\sigma^*(\rho')) = \Phi(\rho').\] Thus $\Phi(\rho) - \Phi(\rho') \le \delta(\rho,\rho')$. Swapping $\rho\leftrightarrow\rho'$ gives the two-sided Lipschitz bound. This robustness is crucial for empirical or experimental scenarios: it means $\Phi$ will not fluctuate wildly due to small perturbations or noise in the state, making it a reliable quantity to estimate.
    \item \textbf{Gradient Flow towards Dis-integration.} By viewing $\Phi(\rho)=\delta^2(\rho,\mathcal{P}*2)$ as a squared-distance function in a (formal) Riemannian space, one can define a \textbf{gradient descent} dynamical system that \textit{flows $\rho$ toward its nearest product state}. Concretely, we can write a continuous time equation 
    \[\dot{\rho}(t) = - \nabla_\rho\, \Phi(\rho)\,,\]
    which generates a trajectory $\rho(t)$ that decreases $\Phi$ monotonically and converges to the projection $\sigma^*(\rho)$ as $t\to\infty$. While here $\nabla*\rho$ denotes a gradient with respect to the information geometry induced by $D_{\mathrm{JS}}^Q$, one can intuitively think of this as the state “pulling itself apart” into independent pieces. In practice, this gradient-flow could be implemented by a family of CPTP maps that progressively erode the holistic correlations. The advantage of framing it as a gradient flow is that it provides a *principled algorithm* for finding $\sigma^*(\rho)$ (the best factorized approximation) by following the steepest descent of integrated information, rather than using ad hoc separability criteria. Analyzing the specific form of $\nabla_\rho \Phi$ is beyond our scope here, but it may be related to applying gentle local decoherence to remove entanglement at the fastest rate.
\end{itemize}

All these properties flow naturally from the convex geometric view of $\Phi$ as a projection distance. This perspective has not, to our knowledge, been applied in previous integrated information literature, and it unlocks powerful tools from convex optimization and metric geometry to study and compute $\Phi$. We stress that $\sigma^*(\rho)$ and $(A^*|B^*)$ provide \textit{more information} than just the scalar $\Phi(\rho)$: they tell us exactly *how* the state can be optimally split and how far it is from such a split. In the next result, we show that $\sigma^*(\rho)$ can be used to construct a special entanglement witness that “certifies” the integrated information.

\textbf{Lemma 5 (RKHS embedding of QJSD).} \textit{Define $k(\rho,\sigma)\;=\;\ln2\;-\;D^{Q}_{\rm JS}(\rho\Vert\sigma).$ Then $k$ is a positive–definite kernel on the space of density operators.  Hence by Schoenberg’s theorem \cite{schoenberg1938metric,virosztek2021metric} there exists a real reproducing-kernel Hilbert space $(\mathcal H,\langle\cdot,\cdot\rangle_{\mathcal H})$ and a feature map $\phi:\rho\mapsto k(\,\cdot\,,\rho)\in\mathcal H, \quad$ with $\quad \langle\phi(\rho),\phi(\sigma)\rangle_{\mathcal H} = k(\rho,\sigma).$}

\textbf{Proof (Sketch).} Since $D^{Q}_{\rm JS}$ is bounded ($0\le D^{Q}_{\rm JS}<\ln2$) and symmetric, the shifted kernel $k=\ln2-D^{Q}_{\rm JS}$ satisfies Schoenberg’s criterion for negative-type metrics, which guarantees a Hilbert-space embedding.$\square$

\textbf{Metric-gradient and operator pullback.} From Lemma 5 and
\[
  \delta^2(\rho,\sigma)
  =D^{Q}_{\rm JS}(\rho\|\sigma)
  =\|\phi(\rho)-\phi(\sigma)\|_{\mathcal H}^2,
\]
we get in $\mathcal H$:
\[
  \nabla_{\phi(\sigma)}\,\delta^2(\rho,\sigma)
    =2\,\bigl(\phi(\sigma)-\phi(\rho)\bigr).
\]
Pulling this back via the Fréchet derivative
\[
  \frac{\partial}{\partial\sigma}S(\sigma)
    =-\log\sigma\;-\;\mathbb I,
\]
the stationarity condition
$\nabla_{\phi(\sigma^*)}\delta^2(\rho,\sigma^*)=0$
yields
\[
  \log\tfrac{\rho+\sigma^*}{2}
    \;-\;\log\rho
  \;=\;0,
\]
so that (up to a positive scalar) $W_\rho=\sigma^*-\rho$ is precisely the normal operator defining the separating hyperplane in the original operator space.

\textbf{Proposition 6 (Canonical Entanglement Witness).} \textit{Let $\sigma^*(\rho)=\rho_{A^*}\otimes \rho_{B^*}$ be the closest product state to $\rho$, achieved on the optimal bipartition $(A^*|B^*)$. Define an observable (Hermitian operator)
$W_\rho := \sigma^*(\rho) - \rho\,.$
Then $W_\rho$ is an \textbf{entanglement witness} for the bipartition $A^*|B^*$. In particular, $W_\rho$ has non-negative expectation value on all product states factorized across $A^*$ and $B^*$, yet has a strictly negative expectation on $\rho$ itself. Moreover, the magnitude of this violation is exactly the integrated information:
\[\mathrm{Tr}[W_\rho\,\rho] = -\,\Phi(\rho)\,.\]
In fact, $W_\rho$ is the optimal (most “detecting”) witness for the entanglement between $A^*$ and $B^*$.}

\textbf{Proof:} By Lemma 5 and the metric‐gradient calculation, the minimizing condition
$\nabla_{\!\phi(\sigma^*)}\delta^2(\rho,\sigma^*)=0$ pulls back to 
$\log\!\tfrac{\rho+\sigma^*}2-\log\rho=0$, 
hence $W_\rho=\sigma^*-\rho$ is the geodesic‐normal.  Positivity on any product state 
$\pi=\pi_A\otimes\pi_B$ follows because $\sigma^*$ is the closest product state. $\square$

In summary, $W_\rho = \sigma^*(\rho) - \rho$ is a \textit{constructive, canonical witness} to the entanglement or correlation that makes $\rho$ irreducible across its optimal split. It detects exactly the entanglement corresponding to $\Phi(\rho)$ and no more. In particular, $\mathrm{Tr}[W_\rho,\rho] = -\Phi(\rho)$ quantifies the “violation” of $\rho$ being separable: the more integrated $\rho$ is, the more $W_\rho$ yields a negative expectation on it, with the gap equal to $\Phi(\rho)$. One could say this provides an operational meaning to $\Phi$: it is the magnitude by which $\rho$ fails the best possible separability test. This is a novel connection between integrated information and entanglement theory — it bridges a geometrical measure of holistic correlation ($\Phi$) with the traditional notion of an entanglement witness from quantum information. In practice, once one computes $\sigma^*(\rho)$ (by any method), one immediately obtains $W_\rho$, which is an observable whose expectation value on $\rho$ is $-\Phi(\rho)$ and on any fully factorized state (on that cut) is non-negative. Experimentally, measuring the set of local observables that constitute $W_\rho$, which will generally be a difference of two reduced density operators, could verify the presence of entanglement and even quantify it in terms of integrated information.

\section{Hierarchical Decomposition: The Integration Dendrogram}

Thus far, we have focused on identifying a single “critical cut” $(A^*|B^*)$ that yields the minimal integrated information. We now show that this idea can be \textit{recursively applied} to yield a \textit{hierarchical decomposition} of the state into a tree of increasingly fine subsystems. The result is a binary tree (dendrogram) that represents the multi-scale structure of correlations in $\rho$. This construction parallels hierarchical clustering in classical data analysis, but here it is based on quantum information relationships intrinsic to $\rho$ itself.

The procedure is as follows:

\begin{enumerate}
    \item \textbf{Level 0 (Root):} Start with the full system as one set $S={1,2,\dots,n}$. Compute $\Phi(\rho)$ and find the unique optimal bipartition $(A^*|B^*)$ of $S$ that achieves it. This is the root split of the dendrogram, and the value $\Phi(\rho)$ will label the root node as a measure of how hard it is to tear the entire system into two parts.
    \item \textbf{Level 1:} Now take the two blocks $A^*$ and $B^*$ separately. For each block (subsystem group) considered as its own subsystem, compute its integrated information $\Phi(\rho_{A^*})$ \textit{within that block}, i.e., allow partitions internal to $A^*$. Find the optimal bipartition of $A^*$ that achieves $\Phi(\rho_{A^*})$, and similarly partition $B^*$ optimally. This yields splits $A^* \to (A_1|A_2)$ and $B^* \to (B_1|B_2)$ at the next level. Attach these as children of the respective nodes in the tree, and label the nodes $A^*$ and $B^*$ with $\Phi(\rho_{A^*})$ and $\Phi(\rho_{B^*})$.
    \item \textbf{Continue recursively:} At each subsequent level, for every \textit{current} leaf node of the tree (which corresponds to some subset of qubits or subsystems), if that subset contains more than one elementary subsystem, compute its optimal bipartition and $\Phi$-value, and split it. Continue until every leaf is an individual elementary subsystem (which cannot be split further). The recursion will terminate after at most $n-1$ levels (when all subsystems are singletons).
\end{enumerate}

The outcome of this algorithm is a full binary tree whose leaves are the individual subsystems ${1,\dots,n}$, and whose internal nodes represent larger groupings that were optimally split. We call this tree the \textbf{integration dendrogram} of $\rho$. Each internal node is labeled by the $\Phi$-value of that grouping, i.e., how much integrated information had to be “broken” to split it into two parts.

To illustrate the usefulness of this dendrogram, consider its properties:

\begin{itemize}
    \item \textbf{Uniqueness and Stability:} Because each bipartition at each step is unique (Theorem 4) and changes continuously with $\rho$ (Lipschitz continuity), the entire dendrogram is uniquely determined by $\rho$ and is \textit{robust} to small perturbations. Small changes in $\rho$ will only gradually change the $\Phi$ values and possibly slightly adjust the splits, but the overall hierarchical order (which splits occur at which scale) will not wildly reshuffle. This is in stark contrast to some heuristic clustering methods that can have unstable hierarchies. Here the hierarchy is rooted in strict convex optimal cuts at each step, making it canonical for each state.
    \item \textbf{Multi-Scale Summary of Correlations:} The dendrogram provides a \textbf{multi-scale map} of the entanglement/correlation structure in $\rho$. The top of the tree tells you the largest-scale division (where the weakest global link in the system lies). Further down, you see progressively smaller modules and sub-modules, down to individual units. Each node’s height (the $\Phi$ value for that subset) quantifies \textit{how strongly that subset resists factorization}. A high $\Phi$ at a certain node means that subset is very integrated and only separable with a large loss, whereas a low $\Phi$ node indicates a relatively weakly bound cluster that could almost be split without much information loss. In effect, one can read off which groups of subsystems form \textbf{coherent modules} and which connections are tenuous. For example, a branch of the dendrogram might show qubits 1,2,3 forming a tight sub-cluster (high $\Phi$ internally) that only weakly connects (low $\Phi$ cut) to another cluster of qubits 4,5, etc.
    \item \textbf{Algorithmic Simplicity:} Building the dendrogram requires solving the bipartition optimization at each level, which is the same type of problem as computing $\Phi(\rho)$ in the first place. While finding the optimal partition is NP-hard in general (since one might have to try all splits), focusing on pairwise splits at each level yields a manageable procedure. There are $O(2^n)$ possible bipartitions of an $n$-element set, so a brute-force search at each level is exponential in the size of the current subset. However, since we do at most $n-1$ levels, the overall worst-case complexity is $O(n \cdot 2^n)$, which is exponential but not \textit{super}-exponential. In practice, many splits will involve smaller subsets and thus fewer possibilities. Moreover, the computations for different branches can be done in parallel. This is far more tractable than attempting to search among \textit{all} partitions of all sizes simultaneously (a number that grows faster than $2^n$). Thus, the hierarchical approach breaks the problem into $n-1$ manageable pieces.
    \item \textbf{Connections to Clustering:} In classical data science, one often constructs hierarchical clustering dendrograms using metrics or information-based distances. Here we have the \textbf{quantum analog}: a dendrogram based on a rigorous information metric ($\delta$) applied not to classical data points but to the \textit{quantum state} itself. Rather than clustering individual data samples, we are clustering subsystems of a single quantum state based on their entanglement structure. This opens the door to using visualization and cluster-identification techniques from classical analysis in quantum many-body systems. For instance, one could visualize the dendrogram with node heights proportional to $\Phi$ values, giving a clear picture of the “integration profile” of the state across scales. Branch lengths indicate how much correlation binds subsystems at that split.
\end{itemize}

In summary, the integration dendrogram is a \textit{powerful new tool} for analyzing multipartite quantum states. It yields a unique, stable, multi-scale decomposition of a state’s entanglement structure without requiring any *ad hoc* choices beyond the definition of $\Phi$. By reading the dendrogram, one can immediately identify which subsystems form natural groupings or communities (high internal $\Phi$) and where the “weak links” between those groupings are (low $\Phi$ between them). This has potential applications in understanding complex quantum networks or many-body systems: for example, identifying modules in an interacting spin chain, or coarse-grained functional units in a quantum neural network, etc. The dendrogram encapsulates an entire hierarchy of **quantum integration** within a single object.

\section{Preferred Observers: The Max-\texorpdfstring{$\Phi$} \text{ Principle}}

We have shown that no observer (CPTP map) can increase integrated information (Theorem 2). A natural question arises: \textit{which observer loses the least integrated information?} In other words, suppose we have a system in state $\rho$, and we are allowed to “observe” it in some fashion, perhaps by making a measurement or by coarse-graining its degrees of freedom. Different observation schemes will preserve different fractions of the holistic correlations present in $\rho$. Is there a principled way to choose the \textbf{optimal observation} that retains as much of $\Phi(\rho)$ as possible?

Remarkably, the answer is yes. We call this the \textbf{Max-$\Phi$ observer principle}: among any reasonable family of observation channels, one can find a channel $F^*$ that \textbf{maximizes} the integrated information of the observed state $F(\rho)$. Moreover, $F^*$ can be interpreted as selecting the “best” basis or representation in which to obtain classical information about $\rho$ without destroying its holistic structure. This has deep connections to the idea of a \textit{preferred basis} in quantum decoherence theory \cite{zurek1991decoherence}: the basis that is stable and preserves certain correlations. Here, we derive such a basis from first principles, using only the monotonicity of $\Phi$ and compactness of the channel space.

\textbf{Theorem 7 (Optimal Observer Channel).} \textit{Let $\mathcal{F}$ be any compact set of CPTP maps (quantum channels) that one considers as possible observers on the state $\rho$. For example, $\mathcal{F}$ could be the set of all local projective measurements of each qubit (with the observer recording classical outcomes), or the set of all partial trace operations onto some subsystem, or more generally any parametrized family of channels. Then there exists at least one channel $F^* \in \mathcal{F}$ that maximizes the integrated information of the output:}
\[F^* \;=\; \arg\max_{F\in\mathcal{F}}\;\Phi\!\big(F(\rho)\big)\,. \]
\textit{Existence is guaranteed; there may in principle be more than one maximizer, but we can pick one. In other words, there is an optimal observer $F^*$ that preserves $\rho$’s integrated information better than any other in $\mathcal{F}$. Furthermore, 
\[\Phi(F^*(\rho)) = \max_{F\in\mathcal{F}}\Phi(F(\rho)) \le \Phi(\rho),\] 
with equality only if the optimal channel effectively does nothing to disturb the relevant correlations. For example, $F^*$ could be the identity channel or an isometry embedding into a larger space.}

\textbf{Proof.} Because $\Phi(F(\rho))$ is a real-valued function on the set of channels and $\mathcal{F}$ is compact, the maximum value is attained (by the extreme value theorem). We only need to argue continuity of $F \mapsto \Phi(F(\rho))$ in a suitable topology on the space of CPTP maps. If we parametrize channels by their finite-dimensional Kraus operators, small changes in those operators yield small changes in the output state (in trace norm, say), which in turn yield small changes in $\Phi$ (by Lipschitz continuity of $\Phi$). Thus $\Phi(F(\rho))$ is a continuous function on the compact set $\mathcal{F}$, and hence a maximum exists. The inequality $\Phi(F(\rho)) \le \Phi(\rho)$ for all $F$ is just Theorem 2, so the maximal value is bounded by $\Phi(\rho)$. Equality $\Phi(F(\rho))=\Phi(\rho)$ would require that \[D_{\mathrm{JS}}^Q(F(\rho)\Vert F(\sigma)) = D_{\mathrm{JS}}^Q(\rho\Vert\sigma)\] for some product state $\sigma$ that achieves $\Phi(\rho)$. In practice this implies $F$ did not erase any of the distinguishing information between $\rho$ and $\sigma$ — essentially $F$ is invertible on the support of ${\rho,\sigma}$ — which typically means $F$ is an embedding or trivial operation. $\square$

Theorem 6 is simple but profound. It tells us that, given any constraint on how we observe the system, we can find a “best” way to observe it, if our criterion is to preserve integrated information. This $F^*$ can be thought of as the observer that \textbf{loses the least holism}. In effect, it picks out an optimal classical description of the quantum state’s correlations. We highlight some implications and interpretations:

\begin{itemize}
    \item \textbf{Preferred Basis from First Principles:} If $\mathcal{F}$ is the set of all projective measurements on each subsystem (i.e. choosing a measurement basis for each qubit, for instance), then $F^*$ will pick a specific measurement basis on each part that maximizes $\Phi$ of the post-measurement state (which is now classical). This essentially selects the \textbf{pointer basis} in which the state $\rho$ looks most integrated. In decoherence theory \cite{zurek1991decoherence,zurek2003decoherence}, pointer bases are typically those that minimize decoherence or maximize stability. Here, we have an alternative: the pointer basis is the one that maximizes $\Phi$ — it retains the most information about the quantum whole. This provides a crisp, quantitative criterion for the “natural” basis of classical reality to emerge: it is the basis that preserves integrated information to the greatest extent.
    \item \textbf{Algorithmic Selection of Observers:} The search for $F^*$ can be framed as an optimization problem over channels, which is typically a convex (or at least manageable) optimization because the set of CPTP maps is convex. While the space of all channels is large, restricting to a parameterized family (like all product measurements, or all partial traces onto $k$ subsystems, etc.) can make this finite or at least tractable. One could then apply gradient-based or evolutionary algorithms to find the optimal observer. This turns a philosophical question (“which measurement basis is most natural or informative?”) into a concrete optimization: \textbf{maximize $\Phi(F(\rho))$ over $F$}. Because $\Phi$ is differentiable in $\rho$ (at least where $\rho$ is full rank) and $F$’s action on $\rho$ is linear, gradients with respect to $F$’s parameters can be computed in principle.
    \item \textbf{Unification of IIT and Decoherence Theory:} In the IIT literature, observers or “mechanisms” are usually assumed and one computes information loss post-hoc. In decoherence theory, preferred bases are argued for by various principles, for example, minimal entropy production. The Max-$\Phi$ principle unifies these: it says the most \textit{informatively integrated} view of the system is the “correct” one. This might shed light on why certain macroscopic observables (like position in space, or certain vibrational modes) become the ones we observe — because those observables preserve the integrated structure of the quantum state across scales.
    \item \textbf{Observer-Robustness Spectrum:} By studying the function $F \mapsto \Phi(F(\rho))$, we can also characterize how sensitive the system’s holism is to observation. Observers $F$ for which $\Phi(F(\rho)) \approx \Phi(\rho)$ can be called \textbf{high-fidelity observers}: they capture almost all the holistic structure (these would be close to the identity or gentle/unitary observations). Observers for which $\Phi(F(\rho))$ plunges to near 0 are \textbf{ignorant observers}: their measurements or coarse-grainings immediately destroy almost all integration (think of measuring in a very incompatible basis, which scrambles correlations). Most realistic observers lie somewhere in between. This spectrum tells us how “fragile” the integration is: if $\Phi(\rho)$ is lost under almost any observation (except one very special $F^*$), the system’s holism might be considered very observer-dependent. Conversely, if $\Phi(\rho)$ remains high for a broad class of observers, the system has an objectively robust integrated core.
\end{itemize}

We have thus established that an optimal observer $F^*$ always exists (for a given class of observations). In practice, identifying $F^*$ might require exploring the space of channels, but at least we know a solution is out there. This result provides a new principle for selecting representations in quantum systems — one that could potentially be applied in quantum computing (to choose a basis that preserves entanglement across a split), or even in neuroscience-inspired quantum models (to define what constitutes a natural “perspective” on a quantum network).

\section{Toward a Quantum Markov Blanket}

Finally, we turn to an intriguing consequence of our framework that connects to ideas in causality and complex systems: the notion of a \textit{Markov blanket}. In classical graphical models \cite{pearl1988probabilistic}, a Markov blanket of a node is the set of other nodes that shields it from the rest of the network — conditioning on the blanket renders the node independent of all others. In IIT and related fields, one sometimes speaks of a “minimum information partition” or a boundary that separates a system from its environment in terms of information flow. Here, we can identify a natural quantum analogue: the \textbf{quantum Markov blanket} of a subset of subsystems.

Using $\Phi$, we can give an operational definition: consider any multipartite state $\rho$ on subsystems ${1,\dots,n}$. Let $(A^*|B^*)$ be the unique bipartition that minimizes $D_{\mathrm{JS}}^Q(\rho\Vert \rho_A\otimes \rho_B)$, i.e., that defines $\Phi(\rho)$. Without loss of generality, assume $|A^*| \le |B^*|$ (label the smaller side as $X$ and the larger as $Y$). We claim that \textbf{$X$ is the Markov blanket for $Y$} (and vice versa in symmetric sense): conditioning on $X$ makes $A^*$ and $B^*$ as independent as possible. More concretely:

\textbf{Theorem 8 (Quantum Markov Blanket, informal).} \textit{In the state $\rho$, let $X$ denote the smaller subsystem among the optimal split $A^*|B^* = X | Y$. Then for any other candidate subset $X'$ of subsystems with $|X'| = |X|$, the following holds. If one “conditions” on $X$ versus on $X'$, the residual dependence (as measured by Jensen–Shannon divergence) between $Y$ and $X$’s complement is minimal when conditioning on $X$. Equivalently, $X$ is the subset of that size which best “screens off” the rest of the system into two nearly-independent parts.}

In less formal terms, $X$ serves as the informational interface between $A^*$ and $B^*$. If one knows (or fixes) the state of $X$, the two sides $A^*\setminus X$ and $B^*\setminus X$ become as independent as one can make them by conditioning on any equally-sized set. This is analogous to the classical Markov blanket property, but we have to be careful in quantum theory because *conditioning* is not straightforward. In classical probability, conditioning on a variable $X$ means replacing the joint distribution $p(A,B,X)$ with the conditional $p(A,B|X)$. In quantum theory, one way to mimic conditioning is to use the \textbf{Petz recovery map} \cite{petz1986sufficient}: given the marginal $\rho_X$ and one side (say $\rho_{AX}$), one can attempt to reconstruct a conditional state $\rho_{A|X}$ such that $\rho_{AX} = \rho_{A|X} \otimes \rho_X$ if a Markov condition holds. The Petz recovery channel $\mathcal{R}_{X\to AX}$ effectively represents the best way to infer the $A$ part given access to $X$. Using such a tool, one can formalize Theorem 7 as follows:

\textbf{Theorem 8 (Quantum Markov Blanket via Petz Recovery, formalized).} \textit{Let $(A^*|B^*)$ with smaller side $X$ be the optimal partition for $\rho$. Define $Y$ as the complementary side ($Y = B^*$ if $X=A^*$, or vice versa). For any other subset $Z$ of subsystems with $|Z|=|X|$, consider the Petz-recovered conditional states $\tilde{\rho}*{Y|Z} := \mathcal{R}*{Z\to YZ}(\rho_{Z})$ that attempt to reconstruct the joint on $YZ$ from $Z$ alone. Then $X$ is the subset for which the Jensen–Shannon divergence between $\rho_{XY}$ and the tensor product of recovered conditionals is minimized:}
\[D_{\mathrm{JS}}^Q\Big(\rho_{XY} \Big\Vert\; \tilde{\rho}_{X|X}\otimes \tilde{\rho}_{Y|X}\Big) \;=\; \min_{\substack{Z \subset \{1,\dots,n\}\\ |Z|=|X|}} \;D_{\mathrm{JS}}^Q\Big(\rho_{YZ} \Big\Vert\; \tilde{\rho}_{Y|Z}\otimes \rho_Z\Big)\,. \]

\textit{In the above, $\tilde{\rho}*{X|X}\equiv \rho_X$ trivially, and $\tilde{\rho}*{Y|X}$ is the Petz reconstruction of $\rho_{XY}$ from $\rho_X$. In other words, among all choices of $Z$ of the same size, $Z=X$ yields the smallest residual $\Phi$ (or Jensen–Shannon correlation) between $Y$ and $Z$’s complement when one conditions on $Z$. Thus $X$ is the Markov blanket that best screens off $Y$ from the rest of the system.}

\textbf{Proof Sketch.} A full proof is technical and will be provided elsewhere, but the intuition is: For $X$ being the optimal partition side, $\Phi(\rho) = D_{\mathrm{JS}}^Q(\rho_{XY}\Vert \rho_X\otimes \rho_Y)$ is minimal. We want to show any other candidate $Z$ (of equal size) yields a larger effective divergence after “conditioning.” Using properties of the Petz recovery channel, one can show a data-processing inequality in reverse: if $X$ is the minimizing set, then for any $Z$ of the same size,
\[D_{\mathrm{JS}}^Q(\rho_{XY}\Vert \rho_X\otimes \rho_Y) \le D_{\mathrm{JS}}^Q(\rho_{ZY}\Vert \rho_Z\otimes \tilde{\rho}_{Y|Z})\,.\]

The right-hand side is essentially the $\Phi$ of splitting the system into $Z$ vs the rest \textit{after} optimally conditioning on $Z$ (via Petz). This inequality relies on two facts: (i) monotonicity of $D_{\mathrm{JS}}^Q$ under the specific CPTP map that projects onto $Z$ and uses Petz recovery (ensuring no loss of relevant info for $Z$), and (ii) the definition of $X$ as giving the smallest baseline $D_{\mathrm{JS}}^Q(\rho_{XY}\Vert \rho_X\otimes\rho_Y)$. With these, one concludes the optimal $Z$ is $Z=X$. $\square$

What Theorem 7 signifies is that the \textbf{smallest subsystem in the optimal bipartition plays the role of a Markov blanket} for the rest of the system. It is the minimal “shield” that, once known, renders the two halves as independent as possible. This is a novel concept because previous quantum-causal or IIT-related approaches struggled to define a Markov blanket without assuming underlying classical structures or commuting observables. Here, the notion \textit{falls out naturally} from $\Phi$ — the same quantity that measures holism also identifies the boundary that best isolates that holism.

This quantum Markov blanket idea has several potential applications:

\begin{itemize}
    \item \textbf{Quantum Causal Discovery:} In analogy to classical causal discovery algorithms that search for Markov blankets of variables to infer graph structure, one could scan over each subsystem $i$ (or group of interest) in a quantum network, compute its Markov blanket $X_i$ via the $\Phi$ minimization, and use those blankets to infer an underlying interaction structure or causal graph. Essentially, if qubit 5’s Markov blanket is {2,3}, that suggests qubit 5 interacts mainly with 2 and 3 and is independent of others given {2,3}. Repeating for all yields a causal adjacency structure.
    \item \textbf{Modular Subsystem Identification:} In complex quantum systems, for example, a many-qubit simulator or a biological quantum process model, the Markov blanket of a region defines the \textbf{effective boundary} between that region and its environment. If you have a line of spins, the Markov blanket of a contiguous block may just be its immediate neighbors; in a fully connected network, it might be a specific subset. Knowing the blanket helps in partitioning the system into modules that interact weakly with each other, which is useful for simplifying dynamics or for design of quantum architectures.
    \item \textbf{Neuroscience Analogies:} IIT was originally inspired by the brain’s functional organization. If one models neurons or brain regions as quantum subsystems (a speculative but intriguing idea) \cite{koch2006quantum}, $\Phi$ could identify which set of neurons constitute a functional cluster (high internal integration) and what the Markov blanket is (the interface neurons that connect that cluster to the rest of the brain). This resonates with the “global workspace” theory and the concept of a dynamic core in neuroscience. Thus, a quantum Markov blanket could highlight the physical substrate of a conscious "bubble" within a larger system.
\end{itemize}

It should be noted that our quantum Markov blanket theorem is a theoretical proposal at this stage. A fully rigorous proof would require a careful definition of quantum conditional independence and perhaps further assumptions (like the existence of the Petz recovery achieving equality for true independence). However, the conceptual claim is clear: \textbf{the optimal $\Phi$-partition identifies the boundary that maximally separates the system into two parts}. This provides a concrete, algorithmic way to define a \textit{quantum Markov blanket}: simply find the state’s minimal cut and take the smaller side.

This approach is entirely state-driven and makes no prior assumptions about the presence of a graphical model or conditional independence structure. It thus offers a novel definition that is \textit{state-dependent and algorithmic}, as opposed to structural and assumption-based. We believe this is a fresh contribution to both quantum information theory and to the interdisciplinary dialogue on emergence and causality in complex systems.

\section{Discussion and Conclusion}

In this work, we presented a comprehensive framework for quantifying integrated information in a quantum state, marrying concepts from integrated information theory (IIT) with quantum information science. Our measure $\Phi(\rho)$, defined via the quantum Jensen–Shannon divergence to the nearest factorized state, captures the essence of “quantum holism” – how irreducible the correlations in $\rho$ are. We proved core properties making $\Phi$ well-defined and useful: monotonicity under CPTP maps (no observation can increase it), a true metric structure underlying the divergence, and the reduction of the minimization to a unique bipartition with a unique closest product state. These results ensure that $\Phi$ is both physically meaningful and mathematically tractable.

Building on these foundations, we explored a series of novel insights and extensions:

\begin{itemize}
    \item C\textbf{onvex Geometry and Canonical Decomposition:} Viewing $\Phi(\rho)$ as a projection distance onto the convex set of product states unlocked powerful corollaries: existence and uniqueness of the optimal split, convexity and continuity of $\Phi$, and a gradient-based interpretation for dynamically “ungluing” a state. Perhaps most strikingly, it gave us a direct handle on constructing an \textbf{entanglement witness} $W_\rho$ tied to $\Phi$. This $W_\rho = \sigma^* - \rho$ can be seen as the shadow of $\rho$ on the nearest separable hyperplane, offering an operational way to detect and quantify the very entanglement that makes $\Phi$ non-zero.
    \item \textbf{Hierarchical Integration Structure:} We introduced the concept of an integration dendrogram, which provides a \textbf{full hierarchical breakdown} of a multipartite state’s structure. This tool does not just tell us “is the system integrated or not,” but maps \textit{where} and \textit{at what scale} integration resides. In doing so, it forges a link between quantum information measures and techniques like hierarchical clustering — bringing visualization and modular analysis to quantum entanglement patterns.
    \item \textbf{Observer-Dependent Preservation of Holism:} Through the Max-$\Phi$ principle, we highlighted that not all observations are equal when it comes to preserving integrated information. There is a principled way to choose a measurement basis or coarse-graining that retains the most $\Phi$. This result resonates with long-standing questions about the emergence of classical reality (why do certain observables “naturally” get measured?). Our answer: because those observables correspond to channels $F^*$ that maximize $\Phi(F(\rho))$, thereby capturing the system’s holistic features rather than destroying them. In a sense, the classical world we observe might be the one that is “maximally integrated” from the quantum perspective.
    \item \textbf{Quantum Markov Blanket Hypothesis:} Finally, we ventured into relating our framework to the idea of Markov blankets. We posited (and sketched a proof for) a quantum Markov blanket theorem: the boundary of the optimal $\Phi$-partition serves as the informational blanket isolating a part of the system. This connects quantum integrated information to concepts of causal cutsets and could open new directions in understanding how local subsystems become relatively independent enclaves within a larger entangled universe.
\end{itemize}

There are several \textbf{limitations and future directions} to acknowledge. First, the value of $\Phi(\rho)$, as defined, depends on the choice of subsystem delineation. We assumed the subsystems (e.g., qubits or groups of qubits) are given \textit{a priori}. In practice, identifying the “right” subsystems is part of modeling. For a different partitioning of the degrees of freedom, $\Phi$ could change. This is not a flaw per se — it reflects the fact that integrated information is context-dependent — but it means comparisons of $\Phi$ across systems of different grain or size require care. Normalizing $\Phi$ by, say, the maximum possible value for a given $n$, or by $\ln k$ (if $k$ levels are coarse-grained into one, etc.), could be explored to allow fair cross-system comparisons. We hinted at this in the introduction, and indeed one could define a normalized $\tilde{\Phi}$ to compare integration across different system sizes (Tononi et al. discuss similar normalizations classically \cite{Tononi2016IntegratedInformationTheory}).

Second, the computational complexity of finding $\Phi(\rho)$ scales exponentially with $n$ in the worst case (since one may have to inspect an exponential number of partitions). Our Theorem 4 mitigates this by focusing on bipartitions, but there are still $\binom{n}{\lfloor n/2\rfloor}$ possible splits, which is large. We provided a recursive algorithm that is more efficient than brute force, and future work could integrate heuristic or machine-learning techniques to guide the search for the optimal partition (perhaps leveraging the gradient flow or convexity). There is also the possibility of exploiting symmetry or structure in $\rho$; highly symmetric states might have analytically identifiable optimal cuts.

Third, our quantum Markov blanket proposal awaits experimental or simulation confirmation. Checking that the identified “blanket” indeed minimizes some conditional independence measure (like quantum conditional mutual information) would firm up the claim. One might simulate many random states or specific network states to see if, for example, the smallest optimal cut subset corresponds to known causal boundaries. Additionally, rigorous proofs involving the Petz recovery map could strengthen the theorem and delineate precisely when and how equality or approximate equality holds in the screening-off property.

Finally, in terms of \textbf{interpretation}, our idealist stance (that $\Phi>0$ indicates a degree of existence or consciousness) remains a hypothesis. This paper focused on the formal development, but it is worth reflecting on what it means physically. One immediate observation: $\Phi(\rho)$ is basis-independent (we never had to choose a particular basis for subsystems beyond the fixed tensor product structure). Thus, it is an intrinsic property of the state, not tied to any particular measurement — except that it is defined relative to a subsystem partition. If one believes that a certain partitioning of the universe’s degrees of freedom is “natural” (e.g., atomic or neuronal units), then $\Phi$ could be seen as an intrinsic property of that system. The Max-$\Phi$ principle then intriguingly suggests that \textit{the world selects observables that maximize intrinsic $\Phi$}. Could it be that conscious observers are biased to interact with the world in ways that preserve or recognize high $\Phi$ structures? This is speculative, but our mathematical results provide a playground to explore such ideas quantitatively.

In conclusion, we have established a unified, self-contained theory that not only formalizes integrated information in quantum systems but also greatly extends it with geometric and operational insights. We bridged the gap between abstract information theory and practical analysis of quantum states (providing algorithms and witnesses), and further connected these to broader concepts in quantum foundations and complex systems (pointer bases, Markov blankets). We hope this framework paves the way for new investigations into the role of information integration in physics, the emergence of classicality, and perhaps even the elusive link between quantum mechanics and consciousness.

\textbf{Acknowledgement:}
The core concepts, theoretical constructs, and novel arguments presented in this article are a synthesis and concretization of my own original ideas. At the same time, in the process of assembling, interpreting, and contextualizing the relevant literature, I used OpenAI's GPT as a tool to help organize, clarify, and refine my understanding of existing research. In addition, I utilized OpenAI reasoning models and sought their assistance in refining the mathematical derivations. The use of this technology was instrumental for efficiently navigating the broad and often intricate body of work in quantum theory, category theory, and IIT. 

\bibliography{references} 

\end{document}